\documentclass[12pt,preprint]{aastex}
\def\lsim{\lower.5ex\hbox{$\; \buildrel < \over \sim \;$}}
\def\gsim{\lower.5ex\hbox{$\; \buildrel > \over \sim \;$}}

\begin{document}

\title{Quasi-Periodic Oscillations in Numerical Simulation of Accretion Flows Around Black Holes}

\author{Sandip K. Chakrabarti$^{1,2}$, K. Acharyya$^3$ and D. Molteni$^4$}

\affil{$^1$ S.N. Bose National Centre for Basic Sciences, JD-Block, Salt Lake, Kolkata 700098,\\ India\\
$^2$ Also at Centre for Space Physics, P-61 Southend Gardens, Kolkata, 700084, India\\
$^3$ Centre for Space Physics, P-61 Southend Garden, Kolkata, 700084, India\\
$^4$ Dipartimento di Fisica e Tecnologie Relative, Viale delle Scienze, 90128 Palermo Italy\\
e-mail: chakraba@bose.res.in, space\_phys@vsnl.com, molteni@unipa.it}

\begin{abstract}

We present results of several numerical simulations of two dimensional 
axi-symmetric accretion flows around black holes using the Smoothed Particle Hydrodynamics (SPH). 
We consider both stellar black holes and as well as super-massive black holes.
We assume bremsstrahlung to be the only source of cooling as it is simpler to
implement in numerical simulations. We observe that due to both radial and vertical 
oscillation of shock waves in the accretion flow, the luminosity and average thermal energy 
content of the inner disk exhibit very interesting behaviors. When power 
density spectra are taken, quasi-periodic oscillations are seen at a few 
Hz and also occasionally at hundreds of Hz for stellar black holes. 
For super-massive black holes, the time-scale of the oscillations
ranges from hours to weeks. The power density spectra have usual flat 
top behavior with average {\it rms} amplitude a few percent 
and a broken power-law behavior. The break frequency is generally found to be
close to the QPO frequency as seen in the observed power spectra of black holes.

\end{abstract}


\noindent (Submitted to Astrophysical Journal)

\section{Introduction}

Galactic black holes are known to exhibit quasi-periodic oscillations (QPOs). 
Today, there are almost a dozen of confirmed stellar mass black hole candidates for which 
QPOs are regularly observed, some at normal frequency ($0.1-10$Hz) and some with
frequency at hundreds of Hz. Molteni, Sponholz and Chakrabarti (1996, hereafter 
referred to as MSC96) 
and Ryu, Chakrabarti, and Molteni (1997) suggested that shock oscillations
might be responsible for these observed QPOs. Subsequently, Chakrabarti \& Manickam (2000)
and Rao, Naik, Vadawale, and Chakrabarti (2000) showed that indeed, the soft X-ray
does not participate in the oscillation, but the hard X-ray, which is supposed to 
be emitted from the post-shock region at the inner edge of the disk, do participate 
in the oscillations, thereby vindicating the original claim that shock oscillation could 
really be the reason of most of the QPOs observed. It has been shown earlier that the
outflows from the post-shock region depend on the shock strength, i.e., the spectral 
states (Chakrabarti, 1999). This has also been tested by various observations (e.g., 
Dhawan et al., 2000). 

In the present paper, we make an effort reproduce the observed power density spectra 
of black hole candidates by simply studying realistic and time-dependent 
low angular momentum flows close to a black hole as a function of flow parameters. For the
sake of simplicity, we use bremsstrahlung cooling only, but increase the accretion rate
so that the cooling is more effective (as in the more realistic case of Comptonization, 
for instance). In the past, enhanced cooling has been modeled by power-law
cooling process (e.g., Langer, Chanmugam \& Shaviv, 1982; MSC96), but we stick to 
bremsstralung only. We find that the flow oscillates vertically due to interaction of the 
inflow and the outflow (e.g., Molteni et al. 2001), and when the parameters are `right', even the 
shocks oscillate, both {\it in radial and in vertical} directions.  As a result of these
oscillations which cause local density and temperature fluctuations,
the cooling rate of the flow can be strongly time-dependent. We made light-curves with
the average luminosity as a function of time and studied their power density spectra 
(PDS). We present only a few illustrative cases to demonstrate that these PDSs 
generally resemble those observed during hard states of black hole candidates.

Considering that an axisymmetric shock fundamentally separates a flow into a fast 
moving (lower density) supersonic and a slow moving (higher density) sub-sonic regions, 
it should not be surprising that its oscillation frequency would lie at or near 
the break frequency. Of course, shocks are not infinitesimally thin. Neither it is true that
the shocks exhibit only a single mode of oscillation. Thus a broad and often multiple peaks
are expected. This is precisely what we see. What is more, since the inner sonic point 
also separates the flow into two parts, namely, the disk-like behaviour and the free-fall
behavior, there is often a weak peak in the PDS at hundreds of Hz, perhaps corresponding 
to oscillation of this inner region.  We shall demonstrate these as well. 
PDS of numerically simulated light curves also has a flat top up to a break frequency, and with 
an {\it  rms} amplitude ranging of a few percent depending on the accretion rate of the flow.

In the next Section, we present the basic set-up of our numerical simulation.
In Section 3, we present some results of the simulation for galactic and extra-galactic 
black holes and analyse them to demonstrate shock oscillations. From the PDS of the 
`simulated' light curves, we show that QPOs are observed and the PDS has the familiar 
flat top shape with a break. Finally in Section 4, we draw our conclusions.

\section{Setup for the Numerical Simulations} 

We are interested to simulate the behaviour of axisymmetric, inviscid flow
in presence of cooling effects. We chose the specific angular momentum to be `low', 
namely, somewhat smaller compared to the Keplerian value. The motivation for this
stems from the fact that accretion processes into a black hole is necessarily 
transonic and the flow has to be sub-Keplerian near the horizon. Furthermore,
the flow can have a substantial amount of sub-Keplerian matter itself because of
accretion of winds. Recent observations do indicate presence of both the Keplerian 
and sub-Keplerian components (Smith, Heindl, and Swank, 2002).

We choose the specific angular momentum $\lambda$ and energy ${\cal E}$ 
of the injected flow in a way such that shocks may form in the flow 
(Chakrabarti, 1989; MSC96) although in the present circumstance, 
the shocks need not be standing and has a much wider parameter 
space (in the sense that the presence of two sonic points 
could be enough to produce these shocks as the fulfillment of the
Rankine-Hugoniot relation is no longer a necessity). We use the Smoothed Particle 
Hydrodynamics (SPH) to do the simulations as in MSC96, but use both the halves of the flow 
to inject matter. The code has been tested for its accuracy against 
theoretical solutions already (MSC96) and we do not repeat them here. 
Needless to emphasize that the `pseudo'-particles have been chosen 
to be `toroidal' in shape and they preserve angular momentum very 
accurately (Molteni, Ryu, and Chakrabarti, 1996). In the long run, 
energy would dissipate, but is found to be negligible compared to our 
main effect here, namely the bremsstrahlung loss. In MSC96, 
suggestion was made that the radial oscillations of the shock may 
take place when the infall time in the post-shock region is comparable to the 
cooling time scale. Presently, we concentrate on the solutions
where such oscillations are present since our goal is to obtain
PDS of the light curves and inspect if they exhibit QPOs.

For the injected parameters, we concentrate on both the
supermassive black holes and the stellar mass black holes. Table 1 gives 
the model runs which we report here. There are basically two Groups of 
inputs. In Group A, we consider the black hole to be super-massive ($M=10^8M_\odot$) 
and in Group B, we consider a stellar mass black hole ($M=10M_\odot$). Different
Cases are run with the density of matter in c.g.s. unit at 
the outer boundary of the numerical grid (which is chosen to 
be $50r_g$) as the parameter. In all the model-runs we 
choose the following parameters: the index $\gamma$ in the 
polytropic relation $P \propto \rho^{-\gamma}$ ($P$ is the 
isotropic pressure and $\rho$ is the gas density) is $5/3$,
the outer boundary $r_{out}=50$, the specific angular 
momentum $\lambda=1.75$, injected radial velocity $\vartheta_r=0.126$, 
the sound velocity $a=0.04$ and the vertical height
at $r_{out}=50$ is $h=15$. The vertical height is so chosen that the flow
remains in hydrostatic equilibrium in the vertical direction at the
point of injection $h \sim a r_{out}^{3/2}$. Here, we measure all the 
distances in units of the Schwarzschild radius of the black hole 
$r_g=2GM_{BH}/c^2$, all the velocities in units of the velocity of 
light $c$ and all the masses in units of $M_{BH}$, the mass of the 
black hole. Here $c$ and $G$ are the velocity of light and the
gravitational constant respectively. It is to be noted that the 
marginally stable (lowest possible angular momentum with a stable Keplerian
orbit) angular momentum is $1.83$ in this unit. We give the accretion 
rates (${\dot m}={\dot M}/{\dot M_{Edd}}$) in units of the Eddington rate 
in the third column. In the fourth column, we give the average 
location of the shock and in the fifth column, the range of variation 
of the shock location is given. In the sixth column, we give the 
average number of `pseudo'-particles in each run and in the seventh column,
the range in which the number varies is provided. In the last 
two columns we present the QPO frequency in Hz and the $rms$ amplitude 
in percentage about which we shall discuss in the next section. 
As in MSC96, we choose the Paczy\'nski-Wiita (1980) pseudo-Newtonian potential 
to describe the spacetime around the Schwarzschild black hole. 

\newpage

\centerline{ TABLE 1: Inputs and extracted parameters for the model runs}

\begin{center}
\begin{tabular}{llllllllll}
\hline
\hline
Model & $\rho_{inj}$ (gm/cc) & ${\dot m}$ &  $<X_s>$ & 
$X^{max}_{s} - X^{min}_{s}$ & $<N>$ & $\Delta N$ & $\nu_{QPO}$ (Hz) & $R_{\it rms}$ ($\%$) \\
\hline
\hline
A.1& 0.95e-14 & 24 & 16& 14-18 & 22250 &4500 & $9.15 \times 10^{-8}$ & 11.2 \\
& & &&  & & & $1.83 \times 10^{-6}$ &  \\
A.2& 1.265e-14 & 32 & 13 & 11-15 & 19250 & 2500& $5.34\times 10^{-7}$ &  11.2  \\
& & &  &  &  & & $2.06\times 10^{-6}$ &   \\
A.3& 3e-14 & 77 & 6 & 3-9 & 12875 & 750 &$3.36\times 10^{-7}$ &  11.2 \\
& & &  &  &  & & $1.37\times 10^{-6}$ &  \\
A.4& 5e-14 & 128 & 5 & 4.5-5.5 & 10500 & 400 & $3.57 \times 10^{-7}$ & 3.3 \\
& &  & &  &  & & $6.88 \times 10^{-7}$ &  \\
\hline
B.1 & 3.6e-10 & 0.09 &15.8 & 14.6-17 & 21950 & 700 & $19.34$  & 3.5\\
B.2  &4.5e-10 & 0.115 & 16.25 & 14.5-18 &22050 & 900 & $19.45 $ & 4.0 \\
B.3  & 4.5e-8 & 11.5 & 24 & 21-30 & 33500 & 5000 & $10.22$&  14.0 \\
 &  & & &  & & & $2.67$ &  \\
B.4  & 4.5e-7 & 115.0 &5 & 4.6-5.4 & 10825 & 350 &  $8.32$ & 3.75 \\
 & &  & & & &  &  $3.58$ &  \\
\hline
\hline
\end{tabular}
\end{center}

In order to study the effect of cooling, we introduce the bremsstrahlung 
process. Thus the specific energy is not conserved. As we increase the 
density of the gas, the accretion rate increases and so does the cooling 
rate. The thermal pressure is thus decreased, especially in the post-shock
region. One of the important conditions of a standing shock (formed due 
to the centrifugal force close to the black hole) is that the thermal 
pressure ($P$) plus ram pressure ($\rho \vartheta_r^2$) should be 
continuous across the shock in the steady state. As the density is 
further increased in the post-shock region due to compression, the 
reduction of thermal pressure due to excessive cooling causes the shock 
to generally move closer to the black hole. In our time-dependent cases, 
we generally see this trend in Group A Cases (Table 1). Along with the 
shock location, there is a systematic decrease in frequency as well
since the flow gets cooler and the time-scale of cooling is increased.

In Group B Cases, within our parameter range, occasionally there were stronger 
shocks which oscillated radially and the interaction between the outflow and the 
inflow also produces vertical oscillations. There were
strong turbulences in the post-shock region which contributed to
pushing the shock backward. As a result, the shock location was not seen to 
change monotonically with accretion rate. The location is clearly 
determined by several effects: centrifugal force, thermal pressure, ram pressure,
energy loss, turbulent pressure etc.

In presence of Comptonization, the complexity is expected to be compounded
since the cooling will depend on the soft photon supply (which may
depend on the Keplerian rate  and the location of the inner edge of the Keplerian disk). This 
will be dealt with in a future work. For reference we may add that the light crossing times
of the horizon for the two classes of the black holes are $r_g/c \sim 10^3$s and $10^{-4}$s
respectively. Since a steady shock  may form anywhere between $10$ to $100r_g$,
the QPO time-period may be close to $10^5$ to $10^8$s for supermassive 
black holes, and close to $10^{-2}$s to $10$s for stellar mass black holes.
Since we chose the outer boundary of simulation to be at $r_{out}=50$, the in-fall time is
roughly $r_{out}^{3/2} \sim 350$. In order to trust the results of our simulations, we ran
the code several hundred times of this timescale (typically $T_{run}\sim 50,000-60,000$ or more).
For a black hole of $M=10M_\odot$, this time corresponds to only $5-6$s.
Because we kept the accretion rate and the cooling type to be fixed for a given case, we reproduce only 
a single type of light curve, namely, that of the hard states which may contain QPOs.

\section{Results}

First we show examples of shocks with radial and/or vertical oscillations.
Fig. 1(a-c) shows the locations of the SPH particles (dots) along with the
velocity vectors (arrows) of every fifth particle for clarity. This case corresponds
to the Case A.2. Matter distribution is shown at three different times (in units of $r_g/c$)
illustrating the vertical and radial motions of the shock which
oscillated  between $11$ to $15$ Schwarzschild radii, mostly 
staying close to $r_s\sim 13r_g$. Some matter could be seen bouncing back
from the centrifugal barrier (empty region) near the axis
and forming giant vortices which interact with the inflow. These vortices
push the post-shock region ($r\lsim 14r_g$) alternatively in the
upper and in the lower halves.

In Fig. 2(a-d), we show the variation of the light curve in Cases A.1-A.4 (marked 
by a-d in the Figure). The light curves are  the variations of the
luminosity of bremsstrahlung radiation (in units of $r_g^2 c$) ($r<50$) with time (in seconds). 
As the density is increased, the average thermal energy gradually goes down
due to the presence of enhanced cooling and thus the luminosity also
goes down. The average location of the shock decreases (Table 1).
The average number of pseudo-particles as well as the variation of the 
number of these particles also go down (Table 1). As a result, with 
the increase in accretion rates, the light curves are 
found to be less noisy, and the amplitude of fluctuation is found to be less. 

In Fig. 3(a-d) we show the results of the Fast Fourier Transform of these light curves using FTOOLS
provided by NASA, the same software package which is used to 
analyse observational results. We find remarkable result of familiar
power density spectra (PDS) with a QPO near the break frequency.
The QPO frequencies are presented in Table 1. We find that, perhaps due to the
presence of both the vertical and radial oscillations, there are 
two QPOs; the frequency of the stronger one
gradually increases with the accretion rate, while that of the
weaker one gradually decreases. 
The occurance of QPO frequencies  at or near the break frequency of the PDS
can be understood by the following: According
to our solution, QPOs occur due to oscillations of the shock waves which separate
two `phases' of the fluid --- the preshock flow is supersonic and weakly radiating,
while the post-shock flow is subsonic and strongly radiating. It appears that the former 
region of the flow produces the `flat-top' region in the PDS, while the later region produces
a PDS with a slope $\sim -2$. The rms amplitude of the flat top
region is shown in Table 1. It is generally similar at 11.2\% except 
in (d) when it is only 3.3\%. 

In Fig. 4(a-d), we present similar results as in Fig. 2 for stellar mass black holes
(Cases B.1 - B.4 of Table 1). Here too, the disk becomes cooler with increasing density 
and the nature of the light curve changes according to whether the shocks are strong or not. 
In Fig. 4(c), the parameters are such that a strong shock forms and it heats up the
disk (accretion rate was not large to cool the post-shock region effectively), 
while for parameters in Fig. 4(d), a very weak shock forms. Shock oscillation
produces a large amplitude noise in the B.3 Case (Fig. 4c). In Fig. 5(a-d),
the PDS of these four Cases are shown. As in Figs. 3(a-d), here too QPOs are
observed. For small black holes QPOs occur at a few Hz the QPO is located at or 
near the break frequency. Radial and vertical oscillations often produce 
multiple peaks in this Case as well. Other details could be seen in Table 1.

Occasionally, peaks at a very high frequency could be seen, but that is 
found to be transient. If the entire light curve is broken into smaller
pieces, weak peaks at frequencies $\sim 100-300$Hz could be seen, though they
do not persist. One example of this is shown in Fig. 6(a) where the PDS of the 
average thermal energy up to $T=36000$ is plotted for the Case B.4. In Fig. 6(b) and (c) 
we show the flow pattern at two times which are separated by only $18$ units
($0.0018$s). The shock located at $\sim 6r_g$ shows two distinctly different
shapes at these two times, bending up and down. Oscillation of a density enhancement at $\sim 6r_g$ 
would have a time period of $\sim R 6^{3/2} \times 10^{-4}$s, where $R$, the compression
ratio is $\sim 2 $ for a weak shock. The corresponding frequency is $340$Hz. The 
weak peak at $306\pm 7.93$Hz (Fig. 6a) may be due to this oscillation.
Taking data for much longer time removes this weak peak. Since average thermal energy
of the disk is a measure of Comptonized spectrum, we have taken the PDS of this quantity 
in Fig. 6a. The emitted luminosity is too noisy to show this property.  

\section{Discussion and conclusions}

In earlier papers, such as in MSC95 and in Chakrabarti \& Manickam (2000), we 
suggested that shock oscillations might contribute to observed QPOs. 
In this present work, using PDS of light-curve obtained from time dependent numerical 
simulations of sub-Keplerian flows, we convincingly demonstrated that the 
shock oscillations could truly be the cause of the QPOs in black hole candidates. We showed that the
computed light-curve produces power density spectra with characteristic QPOs located 
at or near the break frequency which separates the flat-top type and power-law type
PDSs. For stellar black holes, we not only reproduce QPOs at 
reasonable frequencies ($2-20$Hz), but also find that very often high frequency QPOs 
(at around $100-300$Hz) are produced. We demonstrated that
this could be due to oscillation at the very inner edge where the flow becomes supersonic.
For Active Galaxies and Quasars, oscillations at very low frequencies ($\sim 10^{-7}-10^{-6}$ Hz)
are expected since the time scales scale linearly with the mass of the black hole.

We have considered only bremsstrahlung effects for illustration purposes. We have verified
that the average thermal energy of the disk, which is perhaps a measure of Comptonized
hard X-ray luminosity in hard states, also exhibits oscillations at similar frequencies. 
Thus, we believe that the result could be relevant for the study of QPOs. 

SKC thanks the support of 
Indian Space Research Organization for a RESPOND project.  Discussions with Prof. A.R. Rao
(TIFR) and Mr. A. Nandi (SNBNCBS) are also acknowledged.

{}

\vfil\eject
\begin {figure}
\plotone{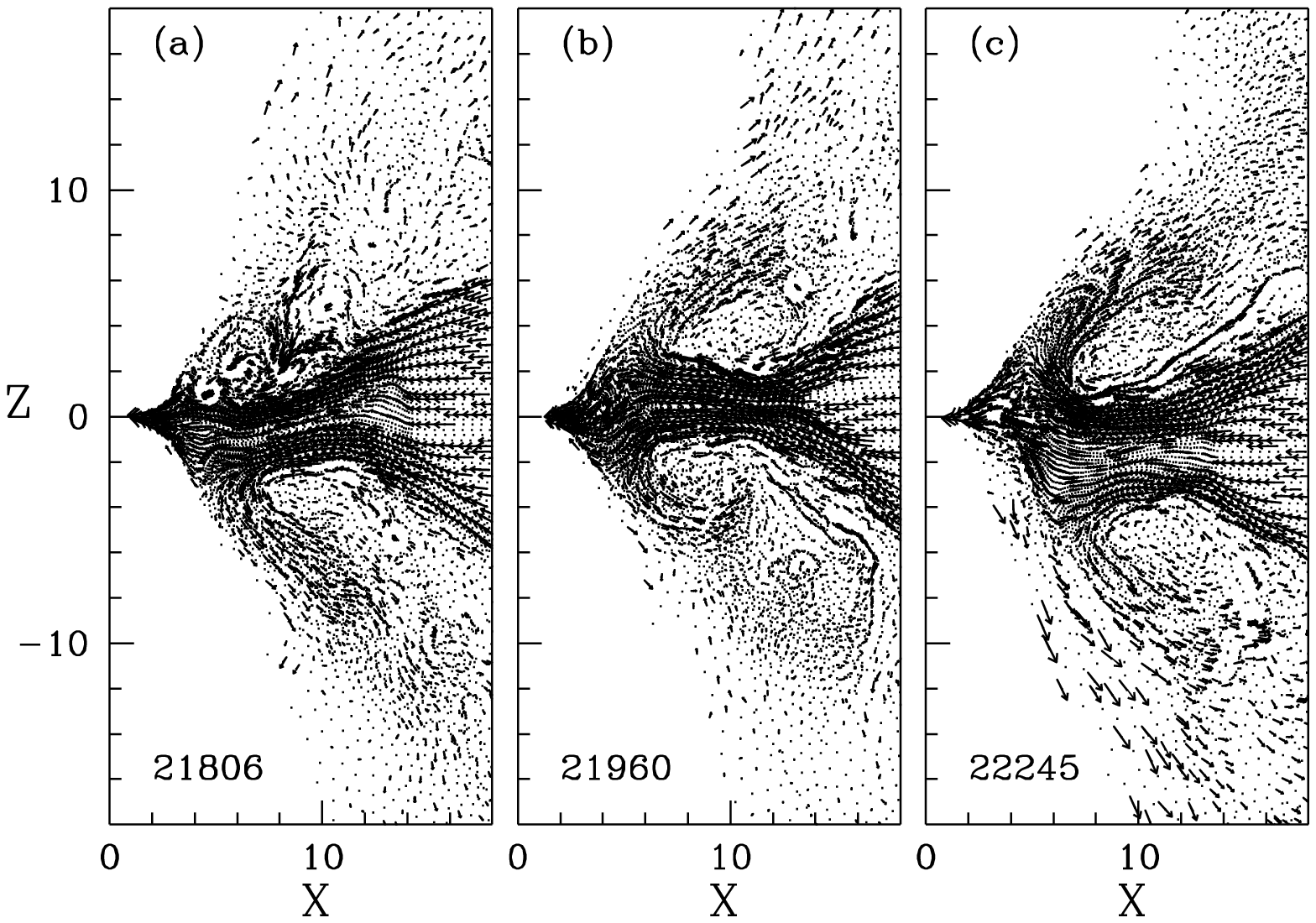}
\caption[] { Snapshots of simulations of accretion disks around a $10^8M_\odot$ black hole
by Smoothed Particle Hydrodynamics. The dots are particle locations and arrows are
drawn for every fifth particle for clarity. Time (in units of $r_g/c$) is marked
on each box. Note the vertical as well as radial oscillation of the accretion shock wave 
located $\sim 13r_g$.}
\end{figure}

\vfil\eject
\begin {figure}
\plotone{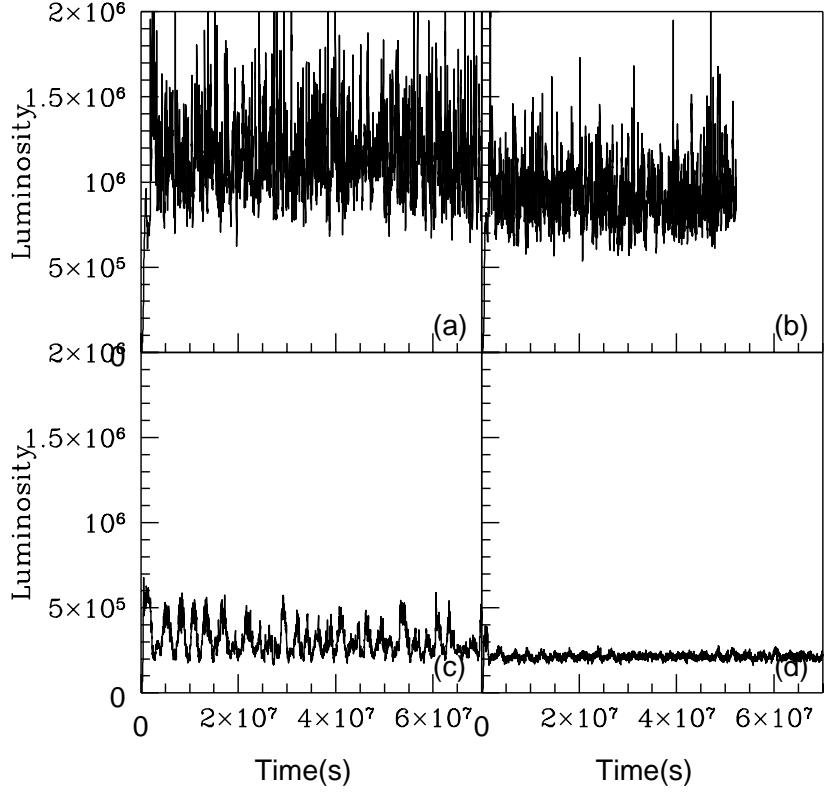}
\caption[] { Total bremsstrahlung luminosity of the accretion flow
(in units of $r_g^2/c$ ) as a function of time (in seconds).
The injected densities are (a) $\rho_{inj}=0.95\times 10^{-14}$gm/s,
(b) $1.265 \times 10^{-14}$gm/s, (c) $3 \times 10^{-14}$gm/s and 
(d) $5 \times 10^{-14}$gm/s respectively. Disk becomes
cooler with increasing accretion rate due to bremsstrahlung
energy loss.}
\end{figure}

\vfil\eject
\begin {figure}
\plotone{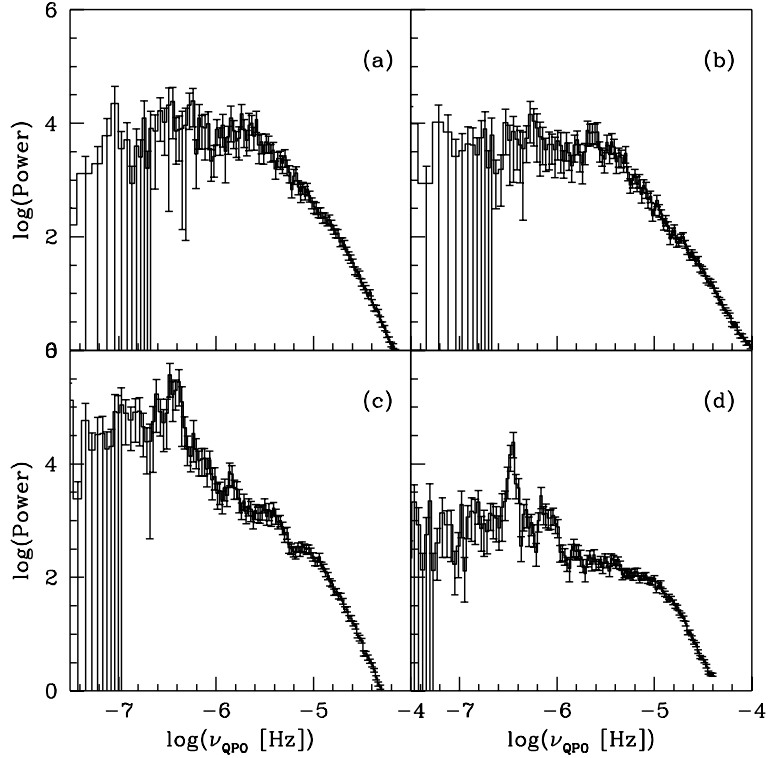}
\caption[] {
Power Density Spectra (PDS) of the four Cases shown in Fig. 2.
Quasi Periodic Oscillation frequencies can be
seen with time-scales of hours to weeks. QPO peaks
are located near break frequencies with flat top
behavior at low-frequency and power-law behavior
at high frequency. See Table 1 for other properties.
}
\end{figure}

\vfil\eject
\begin {figure}
\plotone{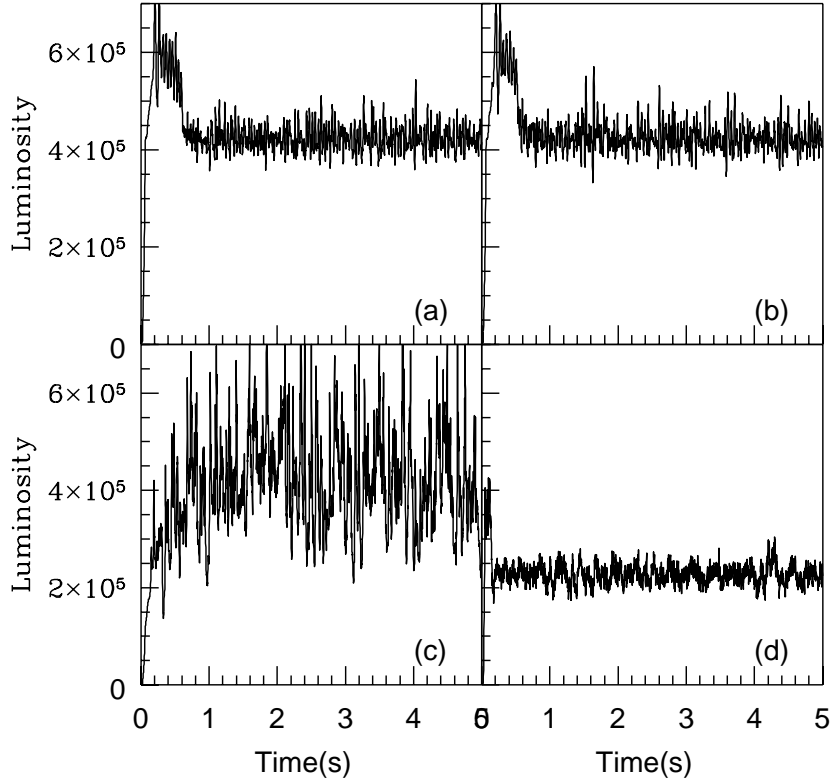}
\caption[] {Total bremsstrahlung luminosity of the accretion flow
(in arbitrary units) as a function of time (in seconds).
The injected densities are (a) $\rho_{inj}=3.6 \times 10^{-10}$gm/s,
(b) $4.5\times 10^{-10}$gm/s, (c) $4.5 \times 10^{-8}$gm/s and 
(d) $4.5 \times 10^{-7}$gm/s respectively. Disk becomes
cooler with increasing accretion rate due to bremsstrahlung
energy loss.
}
\end{figure}

\vfil\eject
\begin {figure}
\plotone{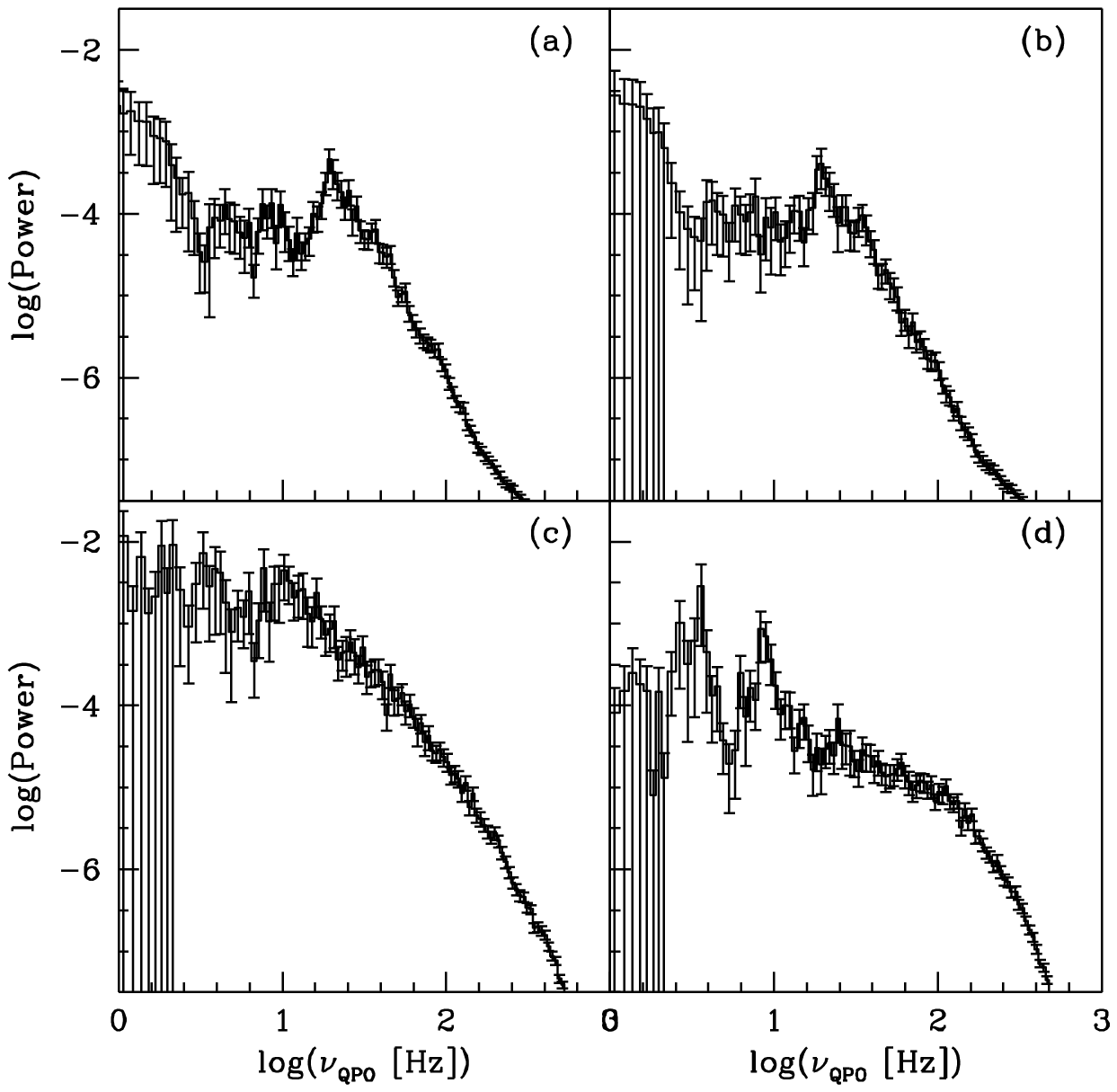}
\caption[] {
Power Density Spectra (PDS) of the four Cases shown in Fig. 4.
Quasi Periodic Oscillation frequencies can be
seen in the range of $2-10$Hz and also
close to several hundred Hzs. QPO peaks
are located near break frequencies with flat top
behavior at low-frequency and power-law behavior
at high frequency. See Table 1 for other properties.
}
\end{figure}

\vfil\eject
\begin {figure}
\plotone{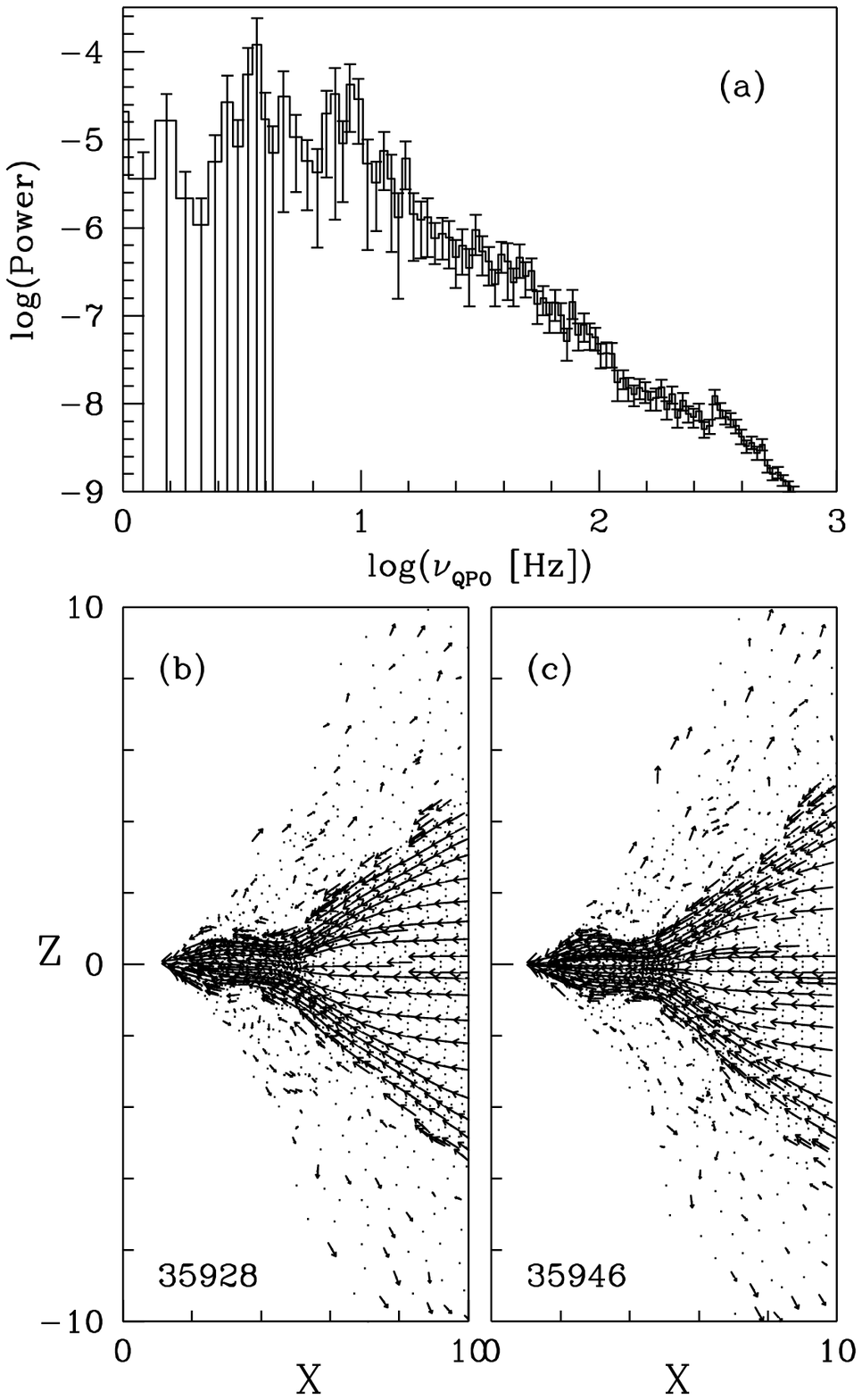}
\caption[] {(a) Power density spectrum of the average thermal energy for Case B.4 with time up to
3.6s. Snapshots of simulations at closely separated times (marked) are
shown in (b-c). Only inner $10$ Schwarzschild radii is shown.
The dots are particle locations and arrows are
drawn for every fifth particle for clarity. Time (in units of $r_g/c$) is marked
on each box. The oscillation of this region causes QPO at $\sim 300$Hz.}
\end{figure}


\begin{thebibliography}{}
\def\ref#1\par{\parshape=2 0in 14.5cm 1cm 13.5cm {#1} \par}
\parskip=0pt
\parindent=0pt

\bibitem[]{}
Chakrabarti S.K. 1989, ApJ, 347, 365

\bibitem[]{}
Chakrabarti S.K. 1999, A\&A 351, 185.

\bibitem[]{}
Chakrabarti S.K., \& Manickam, S.G. 2000, ApJ 531, L41 (CM00)

\bibitem[]{}
Dhawan, V., Mirabel, I. F., \& Rodriguez, L. F. 2000, ApJ 543 373

\bibitem[]{} 
Molteni, D., Acharya, K., Kuznetsov, O., Bisikalo, D. \&  Chakrabarti, S.\ K. 2001, ApJL, 563, L57

\bibitem[]{} 
Molteni, D., Sponholz, H., \& Chakrabarti, S.\ K. 1996, ApJ, 457, 805 

\bibitem[]{} 
Molteni, D., Ryu, D., \& Chakrabarti, S.\ K. 1996, ApJ, 470, 460

\bibitem[]{}
Paczy\'nski, B. \& Wiita, P.J., 1980, A\&A, 88, 23 

\bibitem[]{}
Langer, S.H., Chanmugam, G. \& Shaviv, G., 1982, ApJ, 258, 289

\bibitem[]{}
Rao, A.R., Naik, S., Vadawale, S.V. \& Chakrabarti, S.K. 2000, A\&A 360, L25

\bibitem[]{}
Ryu, D., Chakrabarti, S.K. \& Molteni, D. 1997, ApJ, 474, 378 

\bibitem[]{}
Smith, D. M., Heindl, W. A., \& Swank, J. H., 2002, ApJ, 569, 362

\end{thebibliography}
\end{document}